\documentclass[10pt]{article}
\usepackage{latexsym}
\usepackage{amsfonts}
\usepackage{amsmath}
\usepackage{upgreek}
\usepackage{graphicx}
\usepackage[numbers]{natbib}
\usepackage{subfig}
\usepackage{tikz}
\usepackage{multicol}
\usetikzlibrary{shapes.symbols,shapes.geometric,shadows,arrows.meta}
\tikzset{>={Latex[width=1.5mm,length=2mm]}}
\usepackage{flowchart}\usepackage[paperheight=11.69in,paperwidth=8.26in,left=0.61in,right=0.61in,top=0.61in,bottom=0.61in,headheight=1in]{geometry}
\usepackage[utf8]{inputenc}
\usepackage[T1]{fontenc}
\usepackage{helvet}

\usepackage[font=small,labelfont=bf]{caption}

\newenvironment{Figure}
  {\par\medskip\noindent\minipage{\linewidth}}
  {\endminipage\par\medskip}

\usepackage{tikz}
    \usetikzlibrary{shadows}
\usepackage{tcolorbox}
    \tcbuselibrary{skins}

\definecolor{esoblue}{HTML}{0070C0}
\newcommand{\mytitle}[1]{
    \node[fill=white,
        draw=esoblue,
        line width=0.5pt,
        text width=1.75cm,
        inner sep=8pt,
        xshift=-3.7cm]
    at (frame.north){\bfseries\textcolor{black}{#1}};
}

\newtcolorbox{mybox}[2][]{
    enhanced,
    overlay={\mytitle{#2}},
    borderline={.7pt}{0mm}{esoblue},
    frame hidden,
    arc=0mm,
    sidebyside,
    lefthand width=2.5cm,
    segmentation hidden,
    top=15pt,
    #1
}

\usepackage{xcolor}

\newcommand{\arttitle}[1]{\fontsize{24pt}{32pt}\selectfont \textcolor[HTML]{0070C0}{#1}\par}
\newcommand{\artauth}[1]{\fontsize{12pt}{18pt}\selectfont \textbf{#1}\par}
\newcommand{\artaff}[1]{\fontsize{11pt}{16pt}\selectfont #1 \par}

\newcommand{\TheTitle}{First light of the FIRST visible fibered interferometer upgrade at the Subaru telescope}
\newcommand{\MyName}{Kevin Barjot}
\newcommand{\MyInst}{LESIA, Observatoire de Paris, Université PSL, CNRS, Sorbonne Université, Université de Paris, 5 place Jules Janssen, 92195 Meudon, France}

\begin{document}

\begin{center}
\arttitle{\TheTitle}\par
\artauth{\MyName}\par
\artaff{\MyInst}
\end{center}\par

\vspace{0.5cm}
\begin{multicols}{2}

\noindent Since the mid-nineties, five thousands exoplanets have been discovered, mainly by space missions using indirect detection methods. The existing ground-based 8-m class telescopes and the forthcoming 30-m class telescope offer high angular resolution imaging of planetary system. On the one hand the advantage of such a direct detection is obtaining a spectrum of the planet which permits the study of its atmosphere if it exists and seek for a possible presence of life. On the other hand the caveat is that the atmosphere of the Earth corrugates the light coming from the target thus an (Extreme) Adaptive Optic (ExAO) system is mandatory to correct for these perturbations and perform direct imaging at the diffraction limit regime of the telescope. The pupil masking technique~\cite{tuthill2000aperture} consists in placing a non-redundant mask on the telescope pupil such that each spatial frequency in the image is sampled by a unique pair of sub-pupils. The angular resolution limit is divided by two and the effects of atmospheric perturbations on the fringes are limited. The caveat of this technique is that only a fraction of the flux of the telescope is utilized. It can be coped using the pupil remapping technique~\cite{perrin2006high} which rearrange non-redundantly the sub-pupils using optical fibers and allows to use the flux of the entire pupil. These fibers are single-mode and filter the speckles remaining across each sub-aperture.

\vspace{0.25cm}
{\fontsize{10pt}{10.8pt}\selectfont \textcolor[HTML]{0070C0}{The Fibered Imager foR a Single Telescope}\par}
\noindent The upgraded version of the FIRST instrument~\cite{huby2013} is currently installed on the Subaru Coronagraphic Extreme Adaptive Optic (SCExAO) platform~\cite{scexao2015, vievard2022first}, a second-stage ExAO system providing light corrected from the atmospheric turbulence. As illustrated on the scheme of the two versions of FIRST on the figure~\ref{fig:firstv2Scheme} the telescope pupil is sampled thanks to an array of $250 \, \upmu$m pitched micro-lenses (2) that focus light into single-mode optical fibers (4). This injection is optimized via a segmented mirror (1) paved with 37 hexagonal segments of about $600 \, \upmu$m of circumscribed diameter that can be controlled in piston, tip and tilt. Afterwards, optical delay lines (ODL) (3) were added during the upgrade of the first version to correct for optical fiber length mismatch in order to equalize the optical path differences (OPD) between the incoming beams. The integrated optic (IO) chip (5) combines the light of the sub-pupils pair-wisely as opposed to the first version where the optical fibers were mounted non-redundantly on a V-Groove to make the light beams interfere at the focus of a lens. The chip was tested in laboratory~\cite{barjot2020first, martin2020} prior to its integration in the instrument in 2021. Finally, the chip outputs are imaged on the camera (7) and spectrally dispersed by a grism spectrometer (with a spectral resolution of $4000$ @$700$nm) (6). Polarizations are split by a Wollaston prism in order to avoid contrast loss due to the birefringence of the polarization maintaining fibers and of the chip. FIRSTv2 would ultimately utilize all the sub-pupils to maximize the overall throughput and the spatial frequency coverage of the instrument. 

\begin{Figure}
\centering
\includegraphics[width=\linewidth]{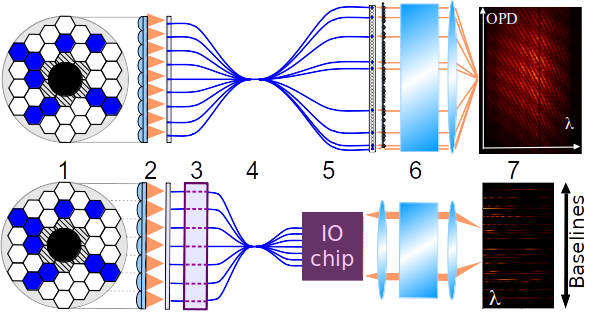}
\captionof{figure}{\small Schematic of the pupil remapping technique implemented in the FIRST instrument. Top: original version. Bottom: version 2 with the IO chip.}
\label{fig:firstv2Scheme}
\end{Figure}
\vspace{0.5cm}

\begin{Figure}
\centering
\includegraphics[width=\linewidth]{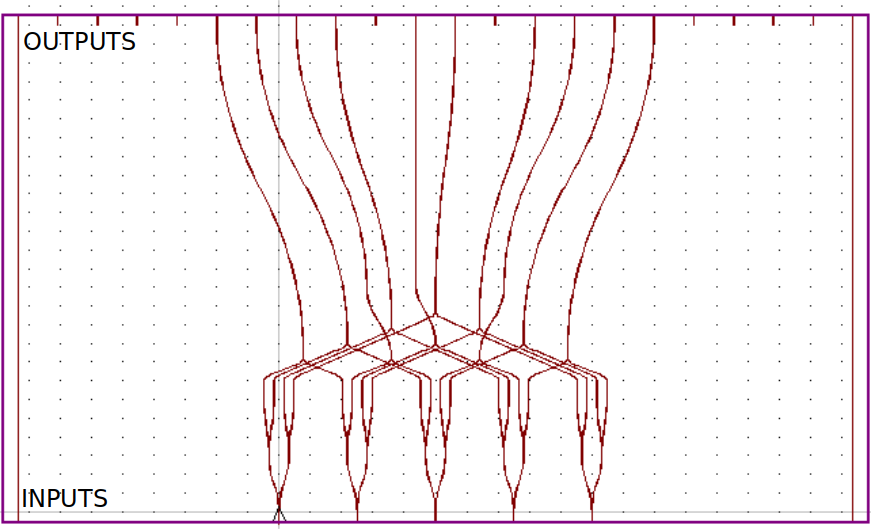}
\captionof{figure}{\small Photonic chip schematic. The five inputs are interferometrically recombined by pairs. Each of the 10 outputs corresponds to one pair.}
\label{fig:photonicScheme}
\end{Figure}
\vspace{0.5cm}

\vspace{0.25cm}
{\fontsize{10pt}{10.8pt}\selectfont \textcolor[HTML]{0070C0}{Data acquisition and processing}\par}
\noindent The key part of the FIRSTv2 instrument is the photonic chip which consists in a block of glass with engraved optical waveguides. As depicted in the schematic of the figure~\ref{fig:photonicScheme} it has five inputs (bottom) that are split in four guides to be recombined thanks to Y-junction couplers giving ten outputs (top). With such a device each of the spatial frequencies that corresponds to each pair is imaged on a few pixels on the camera as a fringe pattern spectrally dispersed. Hence, one frame encodes the interferometric flux for a specific OPD on several spectral channels. In order to obtain the full interferogram the sub-apertures need to be modulated to sample the fringes as a function of the OPD.

Data analysis aims at reconstructing the interferogram from a series of images. A Pixel-To-Visibility Matrix (P2VM)~\cite{millour2004} is calculated prior to observations and is used to fit the fringes and to retrieve the complex coherence terms for each baseline and each spectral channel:
\begin{equation}
	\mu_{nn'}=|V_{nn'}|e^{i\psi_{nn'}}A_nA_{n'}e^{i\Delta\Phi_{nn'}}\label{eq:mu}
\end{equation}

With $|V_{nn'}|$ and $\psi_{nn'}$ the object's complex visibility modulus and phase, $A_n$ the flux of sub-pupil $n$ and $\Delta\Phi_{nn'}$ the differential piston (due to different fiber lengths, Adaptive Optics (AO) residuals and instrumental aberrations).

The phase of this term has two contributions in the context of a binary system observation: the continuum on the whole range of wavelength and the companion on a narrow band range. To obtain the companion contribution $\psi_{compnn'}$ the spectral differential phase is computed subtracting the phase of the continuum $\psi_{contnn'}$ (avoiding the spectral band of the companion) to the phase on the whole spectrum $\psi_{nn'}$, following:
\begin{alignat}{1}
	\upvarphi_{diffnn'}(\sigma) &= \psi_{nn'} + \Delta\Phi_{nn'} - (\psi_{contnn'} + \Delta\Phi_{nn'})\\
	&= \psi_{compnn'}\label{eq:phasefit}
\end{alignat}

Moreover, this variable has the advantage of being self-calibrated by canceling out the differential piston part of the complex coherence terms.

\vspace{0.25cm}
{\fontsize{10pt}{10.8pt}\selectfont \textcolor[HTML]{0070C0}{Lab and on-sky results}\par}
\noindent In the lab two sources were injected on the FIRST testbed to simulate a binary source which spectrum is presented on the bottom left of the figure~\ref{fig:binaryphase}. The wide band continuum and the peak respectively simulate the central star and a companion with a narrow band emission with a contrast of $1.96$ (the companion is $1.96$ times brighter than the central star).

\begin{Figure}
\centering
\includegraphics[width=\linewidth]{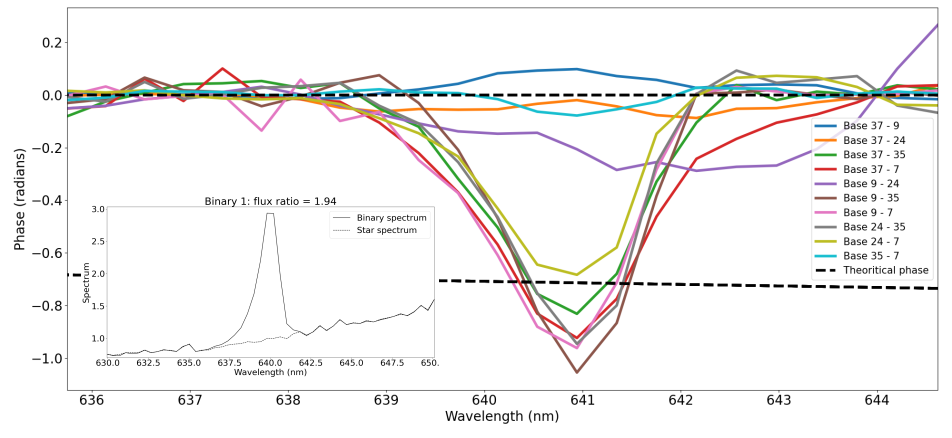}
\captionof{figure}{\small Differential phase measured on the FIRST test bench for a simulated binary system.}
\label{fig:binaryphase}
\end{Figure}
\vspace{0.5cm}

The figure~\ref{fig:binaryphase} is the spectral differential phase of each baseline averaged over $4000$ frames of $300 \,$~ms integration time. The black dashed lines indicate the theoretical phase amplitude expected for such a source. Four of them ($37 - 9$, $37 - 24$, $9 - 24$ and $35 - 7$) show a value of zero because their baseline is orthogonal to the orientation of the binary system. The peak appearing around 641nm for all other baselines show that the companion is detected.

The figure~\ref{fig:onsky} shows the first light of the upgraded FIRST instrument on Spica at the Subaru telescope, in 2021. It shows the fringes of the ten outputs of the photonic chip spectrally dispersed on the vertical axis, for an exposure time of 80ms and a gain factor of 300.

\begin{Figure}
\centering
\includegraphics[width=0.65\linewidth]{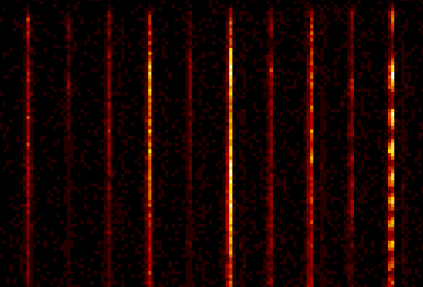}
\captionof{figure}{\small Image of fringes during observations of Spica. Each vertical line corresponds to one baseline. The vertical axis is the direction of the spectral dispersion (from 750nm to 800nm).}
\label{fig:onsky}
\end{Figure}
\vspace{0.5cm}

\vspace{0.25cm}
{\fontsize{10pt}{10.8pt}\selectfont \textcolor[HTML]{0070C0}{Conclusion}\par}
\noindent FIRST is an instrument for high angular and high contrast imaging of binary systems which was upgraded with an integrated optics chip to improve its performances in terms of stability and sensitivity. In 2021 we obtained the first light of the upgraded instrument on the SCExAO platform at the Subaru Telescope. Data processing is still in progress.


\vspace{0.35cm}
{\fontsize{9pt}{9.8pt}
\selectfont \textcolor[HTML]{0070C0}{References}\par 

\begingroup
\renewcommand{\section}[2]{}%

\endgroup
}
\end{multicols}

\vspace*{\fill}
\begin{mybox}{Short CV}
    \includegraphics[scale=.03]{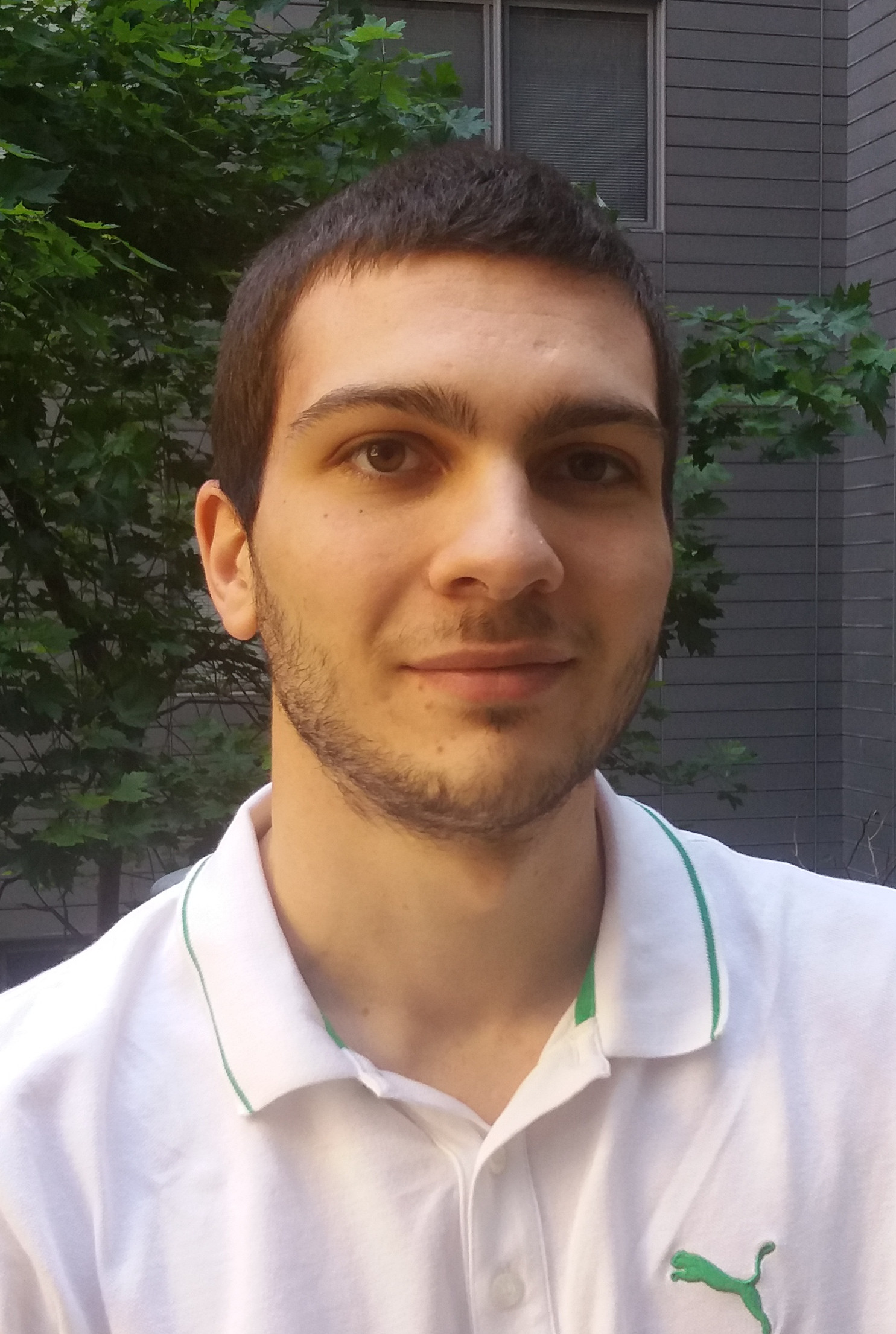}   
    \tcblower
    2017: BSc in Physics, Paris Cité University, France\\
    2019: MSc in Physics and astronomy, Paris Cité University, France\\
    2019-present: PhD in astronomy, Paris Observatory, France
\end{mybox}

\end{document}